\documentclass[prd,twocolumn,floats,floatfix,nofootinbib]{revtex4}
\usepackage[dvips]{graphicx}
\usepackage{graphics}
\usepackage{dcolumn}
\usepackage{amssymb}
\usepackage{bm}
\usepackage{amsmath}
\usepackage{color}

\def\spose#1{\hbox to 0pt{#1\hss}}

\def\lta{\mathrel{\spose{\lower 3pt\hbox{$\mathchar"218$}}
     \raise 2.0pt\hbox{$\mathchar"13C$}}}
\def\gta{\mathrel{\spose{\lower 3pt\hbox{$\mathchar"218$}}
     \raise 2.0pt\hbox{$\mathchar"13E$}}}
\newcommand{\be}{\begin{equation}}
\newcommand{\en}{\end{equation}}
\newcommand{\bea}{\begin{eqnarray}}
\newcommand{\ena}{\end{eqnarray}}

\begin{document}
\title{Bouncing Model in Brane World Theory}

\author{Rodrigo Maier$^{1,2}$, Nelson Pinto-Neto$^{2}$ and Ivano Dami\~ao Soares$^{3}$}

\affiliation{$^{1}$Institute of Cosmology and Gravitation, University of Portsmouth,\\
Dennis Sciama Building, Portsmouth, PO1 3FX, United Kingdom}

\affiliation{$^{2}$ICRA - Centro Brasileiro de
Pesquisas F\'{\i}sicas -- CBPF, \\ Rua Dr. Xavier Sigaud, 150, Urca,
CEP22290-180, Rio de Janeiro, Brazil}

\affiliation{$^{3}$Centro Brasileiro de
Pesquisas F\'{\i}sicas -- CBPF, \\ Rua Dr. Xavier Sigaud, 150, Urca,
CEP22290-180, Rio de Janeiro, Brazil}

\date{\today}

\begin{abstract}

We examine the nonlinear dynamics of a closed Friedmann-Robertson-Walker universe in the
framework of Brane World formalism with a timelike extra dimension. In this scenario, the Friedmann equations contain additional terms arising from the bulk-brane interaction which provide a concrete model for nonsingular bounces in the early phase of the Universe. We construct a nonsingular cosmological scenario sourced with dust, radiation and a cosmological constant. The structure of the phase space shows a nonsingular
orbit with two accelerated phases, separated by a smooth
transition corresponding to a decelerated expansion. Given observational
parameters we connect such phases to a primordial accelerated phase, a soft transition to Friedmann
(where the classical regime is valid), and a graceful exit to a de Sitter
accelerated phase.


\end{abstract}

\maketitle

\section{Introduction}

Although General Relativity is the most successful theory that presently describes
gravitation, it presents some intrinsic crucial problems when we try to construct a
cosmological model in accordance with observational data.
In cosmology, the $\Lambda {\rm CDM}$ model gives us important predictions about
the evolution of the universe and about its current state\cite{mukhanov}. However, let us
assume that the initial conditions of our universe were fixed when the early
universe emerged from the semi-Planckian regime and started its classical
expansion. Evolving back such initial conditions using the Einstein field
equations, we see that our universe is driven towards an initial singularity
where the classical regime is no longer valid\cite{wald}.
\par
Notwithstanding the cosmic censorship conjecture\cite{penrose}, there is no doubt
that General Relativity must be properly corrected or even replaced by a
completely new theory, let us say a quantum theory of gravity. This demand
is in order to solve the issue of the presence of the initial singulariy predicted by
classical General Relativity, either in the formation of a black hole or in the beginning of the
universe. While a full quantum gravity theory remains presently an elusive theoretical problem,
quantum gravity corrections near singularities formed by gravitational collapse have been the
object of much recent research, from quantum cosmology\cite{qc,bojo} to D-brane theory\cite{sahni,2t32,maeda,maartens}.
In the latter scenario, extra dimensions are introduced constituting the bulk space.
In the case of spatially homogeneous and isotropic cosmologies,
the basic resulting distinction between the two approaches lies in the
corrections introduced in the Friedmann Hamiltonian constraint, leading either to modifications
in the kinetic energy terms or to extra potential energy terms. In
both cases we may have bounces in the scale factor corresponding to the
avoidance of a singularity in the models.
In this context, the initial conditions from which our universe has evolved should depend
crucially on the adopted version of the theory to describe the dynamics
around the singularity.
\par
One of the most important characteristics of our universe, supported by
observational data, is its large scale of homogeneity and isotropy. In fact, the
scale of homogeneity and isotropy is empirically well accepted for distances
above $100$ Mpc. Indeed this is the main reason that makes the geometry of
Friedmann-Lemaître-Robertson-Walker (FLRW) a powerful theoretical tool
for the construction of a cosmological scenario\cite{mukhanov}. However, when we consider
a homogeneous and isotropic model filled with baryonic matter, we find
several difficulties when we take into account the primordial state of our
universe. Among such difficulties, we can mention the horizon and flatness
problems\cite{mukhanov}.
\par
As a possible solution to these problems emerged the so called Inflationary
Paradigm\cite{mukhanov,abbott}. Although this fundamental paradigm allows to solve the horizon and flatness
problems, inflationary cosmology does not solve the problem of the initial
singularity. Therefore, nonsingular models from a new theory which provide alternative solutions
to these problems should be strongly considered.
\par
In this paper we adhere to the brane world scenario, where a timelike noncompact
extra dimension is introduced, constituting the bulk space, and all
the matter content of our universe would be trapped on a 4-dimensional
spacetime embedded in the bulk. At low energies General Relativity is
recovered \cite{maeda}, but at high energy scales significant changes are introduced
into the gravitational dynamics and the singularities can be removed \cite{sahni}.
\par
While spacelike extra dimensions theories
have received more attention in the last decades\cite{maartens}, studies involving extra
timelike dimensions have been considered\cite{2t1} despite from the fact that propagating tachyonic modes or negative norm states may arise due to timelike
extra dimensions. These modes have been regarded as problematic once they might
violate causality\cite{dv1} by considering interactions among usual particles.
Issues like the exceedingly small lower bound on
the size of timelike extra dimensions\cite{tvc},
the imaginary self-energy of charged
fermions induced by tachyonic modes -- which seems to cause disappearance of fermions
into nothing -- and the spontaneous decay of stable particles induced
by tachyons with negative energy are major difficulties\cite{dv1}.
Nevertheless, in order to address the cosmological constant problem in Kaluza-Klein theories\cite{2t21}
or reconcile a solution of the hierarchy problem with the cosmological expansion of
the universe\cite{2t22}, timelike extra dimensions have been considered.
On the other hand, it has been shown in \cite{2t31} that the appearance of massless ghosts in an effective four-dimensional theory
can be avoided by considering topological criteria in Kaluza-Klein theories with extra compactified time-like dimensions.
Moreover, avoidance of propagating tachyonic or negative norm states can also be achieved by considering a noncompact
timelike extra dimension\cite{2t32}, which is the case in the model of this paper.
\par
We organize the paper as follows. In section II we present a brief review of the modified Einstein field
equations in the Brane World scenario. In section III, we construct a nonsingular
cosmological scenario sourced with dust, radiation and a cosmological constant.
In section IV, we show that given the observational
parameters, we can connect such phases to a primordial accelerated expansion, a soft transition to Friedmann
(where the classical regime is valid), and a graceful exit to a de Sitter
accelerated phase. As our final remarks, we discuss some of
its possible imprints in the physics of cosmological perturbations.

\section{Field Equations}

For the sake of completeness let us give a brief introduction to braneworld theory,
making explicit the specific assumptions used in
obtaining the dynamics of the model. We rely on references
\cite{sahni,maartens}, and our notation
basically follows \cite{wald}. Let us start with a 4-dim Lorentzian
brane $\Sigma$ with metric $g_{ab}$, embedded in a 5-dim conformally flat bulk $\cal{M}$ with
metric $g_{AB}$. Capital Latin indices range from 0 to 4, small Latin
indices range from 0 to 3. We regard $\Sigma$ as a common boundary of two pieces
${\cal{M}}_{1}$ and ${\cal{M}}_{2}$ of $\cal{M}$ and the metric $g_{ab}$ induced on the brane by the metric of the two pieces should coincide although the extrinsic
curvatures of $\Sigma$ in ${\cal{M}}_{1}$ and ${\cal{M}}_{2}$ are allowed to be different.
The action for the theory has the general form
\begin{eqnarray}
\label{eq4r}
\nonumber
S=\frac{1}{2\kappa^2_{5}}\Big\{\int_{M_{1}}\sqrt{-\epsilon ~^{(5)}g} \Big[^{(5)}R-2\Lambda_{5} +2\kappa^2_{5}L_{5}\Big]d^5x\\
\nonumber
+\int_{M_{2}}\sqrt{-\epsilon ~^{(5)}g} \Big[^{(5)}R-2\Lambda_{5}+2\kappa^2_{5}L_{5}\Big]d^5x\\
\nonumber
+2\epsilon\int_{\Sigma}\sqrt{- ^{(4)}g}K_{2}d^4x-2\epsilon\int_{\Sigma}\sqrt{-^{(4)}g}K_{1}d^4x\Big\}\\
\nonumber
+\frac{1}{2}\int_{\Sigma}\sqrt{-^{(4)}g}\Big(\frac{1}{2\kappa^2_{4}}^{(4)}R-2\sigma\Big)d^4x\\
~~~~~~~~~~~~~~~~+\int_{\Sigma}\sqrt{-^{(4)}g} L_{4}(g_{\alpha\beta},\rho)d^4x.
\end{eqnarray}
In the above $^{(5)}R$ is the Ricci scalar of the Lorentzian 5-dim metric $g_{AB}$
on $\cal{M}$, and $^{(4)}R$ is the scalar curvature of the induced metric $g_{ab}$
on $\Sigma$. The parameter $\sigma$ is denoted the brane tension. The unit vector $n^{A}$ normal to the boundary $\Sigma$ has norm $\epsilon$. If $\epsilon=-1$ the
signature of the bulk space is $(-,-,+,+,+)$, so that the extra dimension is timelike.
The quantity $K=K_{ab}~ g^{ab}$ is the trace of the
symmetric tensor of extrinsic curvature
$K_{ab}= Y_{,a}~^{C}~Y_{,b}~^{D}~{\nabla_{C}}{n_{D}}$, where $Y^{A}(x^a)$ are the
embedding functions of $\Sigma$ in $\cal{M}$\cite{eisenhart}. While
$L_{4} (g_{ab}, \rho)$
is the Lagrangean density of the perfect fluid\cite{taub}(with equation of state $p= \alpha\rho$),
whose dynamics is restricted to the brane $\Sigma$, $L_{5}$ denotes the lagrangian of matter in the bulk. All integrations over the bulk and the
brane are taken with the natural volume elements $\sqrt{-\epsilon~ {^{(5)}}g}~ d^{5}x$ and $\sqrt{- {^{(4)}}g}~ d^{4}x$ respectively. $\kappa_{5}$ and $\kappa_{4}$ are Einstein constants in five and four-dimensions. We use units such that $c=1$.
\par Variations that leave the induced metric on $\Sigma$
intact result in the equations
\begin{eqnarray}
\label{eq6r}
^{(5)}G_{AB}+ \Lambda_5~{^{(5)}}g_{AB}=\kappa^2_5 {^{(5)}}T_{AB},
\end{eqnarray}
while considering arbitrary variations of $g_{AB}$ and taking into account (\ref{eq6r})
we obtain
\begin{eqnarray}
\label{eq7r}
^{(4)}G_{ab}+\epsilon ~\frac{\kappa_{4}^{2}}{\kappa_{5}^{2}}\Big( S^{(1)}_{ab}+S^{(2)}_{ab}\Big )
=\kappa_{4}^{2}\Big(\tau_{ab}-\sigma g_{ab} \Big),
\end{eqnarray}
where $S_{ab} \equiv K_{ab}-K g_{ab}$, and $\tau_{ab}$ is the energy momentum
tensor associated to $L_4$. In the limit
$\kappa_4 \rightarrow \infty$ equation (\ref{eq7r}) reduces to
the Israel-Darmois junction condition\cite{israel}
\begin{eqnarray}
\label{eq9r}
\Big( S^{(1)}_{ab}+S^{(2)}_{ab}\Big )
=\epsilon~ \kappa_{5}^{2}\Big(\tau_{ab}-\sigma g_{ab} \Big)
\end{eqnarray}
We impose the $Z_2$-symmetry\cite{maartens} and use the junction conditions (\ref{eq9r})
to determine the extrinsic curvature on the brane,
\begin{eqnarray}
\label{eq10r}
K_{ab}=-\frac{\epsilon}{2} \kappa_{5}^{2} \Big[(\tau_{ab}-\frac{1}{3}\tau g_{ab})+\frac{\sigma}{3} g_{ab} \Big].
\end{eqnarray}
\par
Now using Gauss equation
\begin{eqnarray}
\label{gauss}
\nonumber
^{(4)}R_{abcd}=~^{(5)}R_{MNRS} Y^{M}_{,a} Y^{N}_{,b} Y^{R}_{,c} Y^{S}_{,d}~~~~~~~~~~\\
+\epsilon \Big(K_{ac}K_{bd}-K_{ad}K_{bc} \Big)
\end{eqnarray}
together with equations (\ref{eq6r})
and (\ref{eq10r}) we arrive at the induced field equations on the brane
\begin{eqnarray}
\label{eq1.2.13}
\nonumber
^{(4)}G_{ab}=-\Lambda_{4}{^{(4)}g}_{ab}+8\pi G_{N}\tau_{ab}+\epsilon\kappa^4_{5}\Pi_{ab}~~~~~~\\
-\epsilon{E}_{ab}+\epsilon
F_{ab}
\end{eqnarray}
where we define
\begin{eqnarray}
\label{eq1.2.14}
\Lambda_{4}~:=\frac{1}{2}\kappa^2_{5}\Big(\frac{\Lambda_{5}}{\kappa^2_5}+\frac{1}{6}\epsilon\kappa^2_{5}\sigma^2\Big),~~~~~~~~~~~~~~~~~~~~~~~~~~~~\\
G_{N}:=\epsilon\frac{\kappa^4_{5}\sigma}{48\pi},~~~~~~~~~~~~~~~~~~~~~~~~~~~~~~~~~~~~~~~~~~~~~~~~\\
\nonumber
\Pi_{ab}:=-\frac{1}{4}\tau_{a}^{c}\tau_{bc} +\frac{1}{12}\tau\tau_{ab}
+\frac{1}{8}{^{(4)}g}_{ab}\tau^{cd}\tau_{cd}~~~~~~~~~~~\\
-\frac{1}{24}\tau^2{^{(4)}g}_{ab},~~~~~~\\
\nonumber
F_{ab}:=\frac{2}{3}\kappa^2_{5}\Big\{\epsilon ~{^{(5)}T}_{BD} Y^B_{,a} Y^D_{,b}~~~~~~~~~~~~~~~~~~~~~~~~~~~~\\+ \Big[{^{(5)}T}_{BD} n^{B} n^{D}-\frac{1}{4}\epsilon~{^{(5)}T}
\Big]{^{(4)}g}_{ab}\Big\},
\end{eqnarray}
$E_{ab}$ is the electric part of the Weyl tensor in the bulk induced in the brane,
$T_{AB}$ is the energy-momentum in the bulk, and
$G_N$ defines the Newton's constant on the brane. For a timelike extra dimension we have that
$\epsilon=-1$ in our conventions implying that $\sigma < 0$ in accordance with observations.
We also remark that the effective 4-dim cosmological constant might be set zero, or made conveniently small,
in the present case of an extra timelike dimension by properly fixing the bulk cosmological constant as
$\Lambda_{5}\simeq \frac{1}{6}\kappa_{5}^{4} ~\sigma^2$.
It is important to note that for a 4-dim brane embedded in a conformally flat empty bulk we have the absence of the Weyl conformal
tensor projection $E_{ab}$ and of $F_{ab}$ in Eq. (\ref{eq1.2.14}).
\par Accordingly, Codazzi's equations imply that
\begin{eqnarray}
\label{eq1.2.17}
\nabla_{a} K-\nabla_{b} K^{b}_{a}=
-\frac{1}{2}\epsilon \kappa^2_{5}\nabla_{b}\tau^{b}_{a},
\end{eqnarray}
resulting in
\begin{eqnarray}
\label{eq1.2.18}
\nabla^{a}{E}_{ab}=\nabla_{b}\tau^{b}_{a}+\kappa^4_{5}\nabla^{a}\Pi_{ab}+\nabla^{a}F_{ab},
\end{eqnarray}
where $\nabla_a$ is the covariant derivative with respect to the induced metric $g_{ab}$.
Equations (\ref{eq1.2.13}) and (\ref{eq1.2.18}) are the dynamical equations of the
gravitational field on the brane.

\section{The Model}

Let us consider a FLRW geometry on the four-dimensional brane embedded
in a five-dimensional deSitter bulk with a timelike extra dimension ($\epsilon=-1$)\cite{sahni}.
Considering comoving coordinates on the brane, the line element is given by
\begin{eqnarray}
\label{eq2}
ds^2=-dt^2+a^2(t)\Big[\frac{1}{1-kr^2}dr^2+r^2d\Omega^2\Big],
\end{eqnarray}
where $a(t)$ is the scale factor, $k$ is the spatial curvature and $d\Omega^2$ is the solid angle.
\par The matter content of the model, restricted to the brane, is given by noninteracting perfect fluids,
namely, dust and radiation, with respective equations of state $p_{dust}=0$, $p_{rad}=\rho_{rad}/3$, and energy momentum tensor
$\tau^{ab}$:=$\tau^{ab}_{~~dust}+\tau^{ab}_{~~rad}$ satisfying $\nabla_{b}\tau^{ab}_{~~dust}=0=\nabla_{b}\tau^{ab}_{~~rad}$.
In this situation we have that
{\small
\begin{eqnarray}
\label{eq2.1}
\nonumber
\Pi_{00}&=& \frac{1}{12} (\rho_{dust}+\rho_{rad}) ^2,\\
\nonumber
\Pi_{ij}&=&\Big[\frac{1}{12}(\rho_{dust}+\rho_{rad})^2 \\
&+&\frac{1}{6}(\rho_{dust}+\rho_{rad}) (p_{dust}+p_{rad})\ \Big] g_{ij},
\end{eqnarray}}
and Codazzi's equations (\ref{eq1.2.17}) imply that $\nabla_{a} \Pi^{a}_{b}=0$, consistent with the
contracted Bianchi's identities in (\ref{eq1.2.13}) and Codazzi's equation (\ref{eq1.2.18}).
The modified Friedmann equations have the first integral
\begin{figure}
\begin{center}
{\includegraphics*[height=7cm,width=8cm]{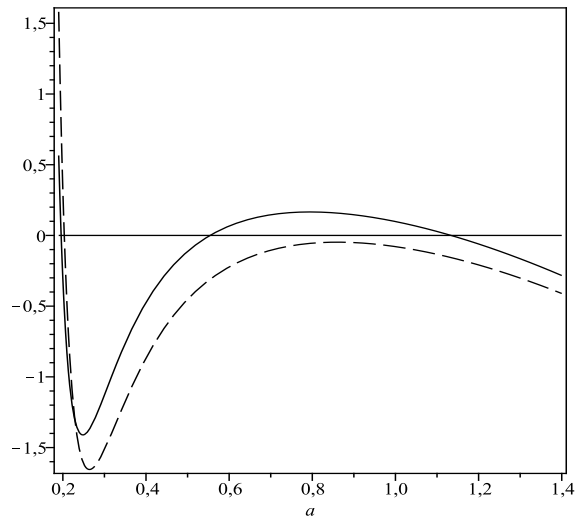}}
\caption{Illustration of the behavior of $V(a)$ for parameters $k=0.8$, $\Lambda_4=1.5$ $E_{rad}=0.1$, and
for $E_{dust}=0.001$ (continuous line) $E_{dust}=0.09$ (dashed line). The increase of the dust content for
a fixed $k$ excludes the presence of perpetually bouncing solutions.}
\label{fig0}
\end{center}
\end{figure}
\begin{figure}
\begin{center}
{\includegraphics*[height=7cm,width=8cm]{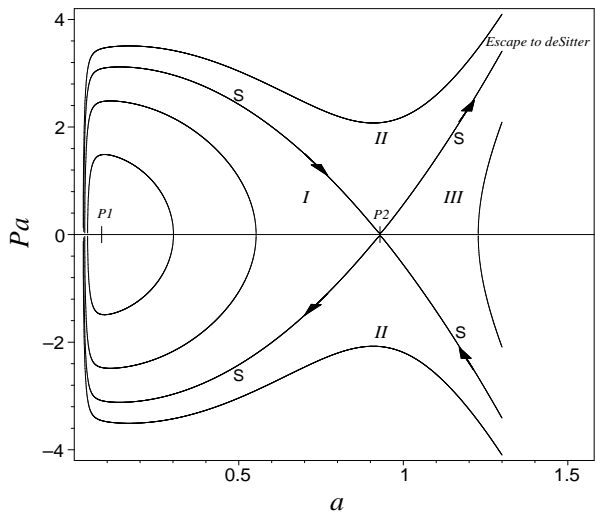}}
\caption{Phase portrait of the dynamics with the critical point $P_1$ (center) and $P_2$ (saddle). Orbits in region II
are solutions of one-bounce universes with a graceful exit to an accelerated (inflationary) phase along the separatrix ${\mathcal S}$.}
\label{fig-plane}
\end{center}
\end{figure}
\begin{eqnarray}
\label{eq4}
\nonumber
H^2&+&\frac{k}{a^2}-\frac{\Lambda_{4}}{3}=\frac{8 \pi G_N}{3}(\rho_{dust}+\rho_{rad})\\
&-&\frac{4 \pi G_N }{{3|\sigma|}}(\rho_{dust}+\rho_{rad})^2~,
\end{eqnarray}
where $H:={\dot{a}}/{a}$ with $\dot{a}\equiv {da}/{dt}$.
It is worth noting that the bounce is solely engendered due to the presence of a timelike
extra dimension which induces the minus sign in the last term of (\ref{eq4}).
By assuming a spacelike extra dimension, we would obtain a plus sign instead
that provides a singular model.
\par
Expressing
\begin{eqnarray}
\label{eq3}
\rho_{dust}=\frac{E_{dust}}{a^3}~,~~~~~\rho_{rad}=\frac{E_{rad}}{a^4},
\end{eqnarray}
where $E_{dust}$ and $E_{rad}$ are constants of motion, the first integral of motion (\ref{eq4})
can be expressed as the Hamiltonian constraint
\begin{eqnarray}
\label{eq3.6}
{\cal{H}} =\frac{p_{a}^2}{2}+V(a)=0,
\end{eqnarray}
where
\begin{eqnarray}
\label{eq3.5}
\nonumber
V(a)=\frac{k}{2}&-&\frac{\Lambda_{4}a^2}{6}-\frac{8 \pi G_N}{6}\Big(\frac{E_{dust}}{a} +\frac{E_{rad}}{a^2}\Big)\\
&+&\frac{8 \pi G_N}{12|\sigma|}\Big(\frac{E_{dust}}{a^2}+\frac{E_{rad}}{a^3}\Big)^2.
\end{eqnarray}
From (\ref{eq3.6}) we derive the dynamical system
\begin{eqnarray}
\label{eq5}
\dot{a}=p_a~,~\dot{p}_a=-\frac{dV}{da}.
\end{eqnarray}
It is worth noting that the last term in the potential (\ref{eq3.5}) acts as an infinite potential barrier and is responsible for
the avoidance of the singularity $a=0$. These potential corrections are equivalent to fluids with negative energy densities. This is in
accordance with the fact that quantum effects can violate the classical energy conditions, and may avoid
curvature singularities where classical general relativity breaks down\cite{np}. Such violations tend to occur
on short scales and/or at high curvatures, which is the case of the present models.
\par The behavior of the potential $V(a)$ is illustrated in Fig. \ref{fig0} for $k=0.8$, $\Lambda_4=1.5$, $E_{rad}=0.1$, and for
$E_{dust}=0.001$ (continuous line) $E_{dust}=0.09$ (dashed line). We can see that the increase of the dust content for
a fixed $k$ excludes the presence of perpetually bouncing solutions by driving the maximum out of the physical space.
For a sufficiently large $E_{rad}$ the potential $V(a)$ presents no local maximum or minimum\cite{maier}.
\par The critical points in the phase space are stationary solutions of (\ref{eq5}), namely,
the points of the phase space $(a=a_{crit},p_a=0)$ corresponding to the zeros of the right-hand-side of (\ref{eq5}).
Here, $a_{crit}$ stands for the real positive roots of $dV/da$. By considering the case of closed geometries ($k>0$),
it is not difficult to verify that, depending on the values of the parameters
$(\Lambda_4,|\sigma|,E_{rad},E_{dust})$, there are at most two critical points associated with one minimum
and one local maximum of $V(a)$. In this case, the minimum of the potential corresponds to a center while
the maximum corresponds to a saddle.
This configuration allows us to
obtain different types of orbits that describes the evolution of universes in this model.
In Fig. 2 we illustrate the phase space portrait of the model for $\Lambda_4=1.5$, $\sigma=6000$, $E_{rad}=0.15$ and $E_{dust}=0.05$,
and for varying $k$. The value of $E_{dust}$ is sufficiently bounded so that $V(a)$ has a well. The critical points
$P_1$ (center) and $P_2$ (saddle) correspond to stable and unstable Einstein universes. Typically the model allows for the
presence of perpetually bouncing universes (periodic orbits in region I), and one-bounce universes (region II).
Region I is bounded by the separatrix $\mathcal{S}$ emerging from the saddle $P_2$. A separatrix also
emerge from $P_2$ towards the deSitter attractor at infinity, defining a graceful exit of orbits in region II to an (inflationary) accelerated phase.
From now on we will restrict ourselves to the case of
closed geometries. In the next section we will exam what kind of orbit would be generated when one considers the
observational values of $(\Lambda_4,E_{rad},E_{dust})$.

\section{Observational Cosmology}
\begin{figure}
\begin{center}
{\includegraphics*[height=20cm,width=8cm]{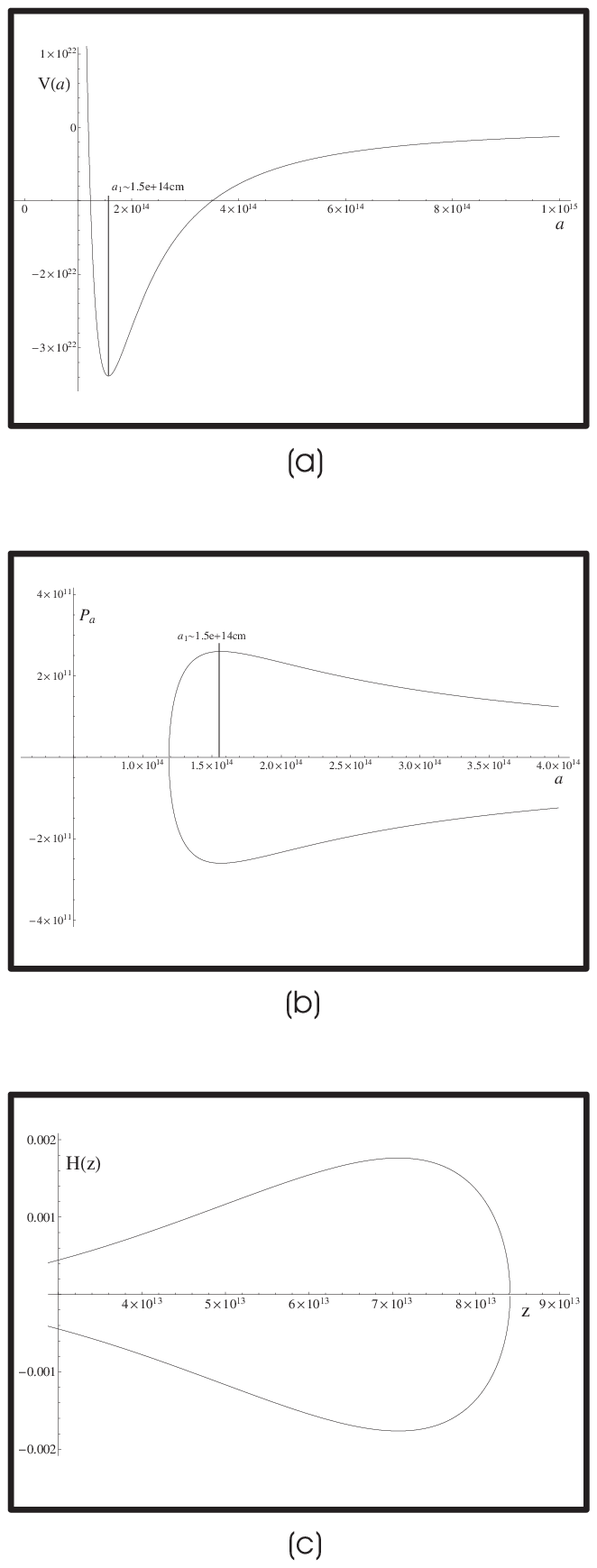}}
\caption{(a) The potential $V(a)$ and the phase space (b) in the region that encompasses the critical point $a_1$, considering the observational values (\ref{eqoc1}) and (\ref{eqocbt}). For $|\sigma|\sim|\sigma|_{min}$, we obtain $a_1\sim10^{14}~{\rm cm}$. Although this primordial accelerated phase does not correspond to usual inflation ($0.27$ being the number of e-folds), it is important to remark that as our universe
has no beginning of time and the cosmological constant
is small, the particle horizon before the bounce
was already bigger than the scales of cosmological
interest.
In (c) we show the behavior of the Hubble factor as function of redshift in a neighborhood of the bounce given the normalization $a_0=1$ and the parameters (\ref{eqrs4}).}
\label{Fig. 1}
\end{center}
\end{figure}
\begin{figure}
\begin{center}
{\includegraphics*[height=20cm,width=8cm]{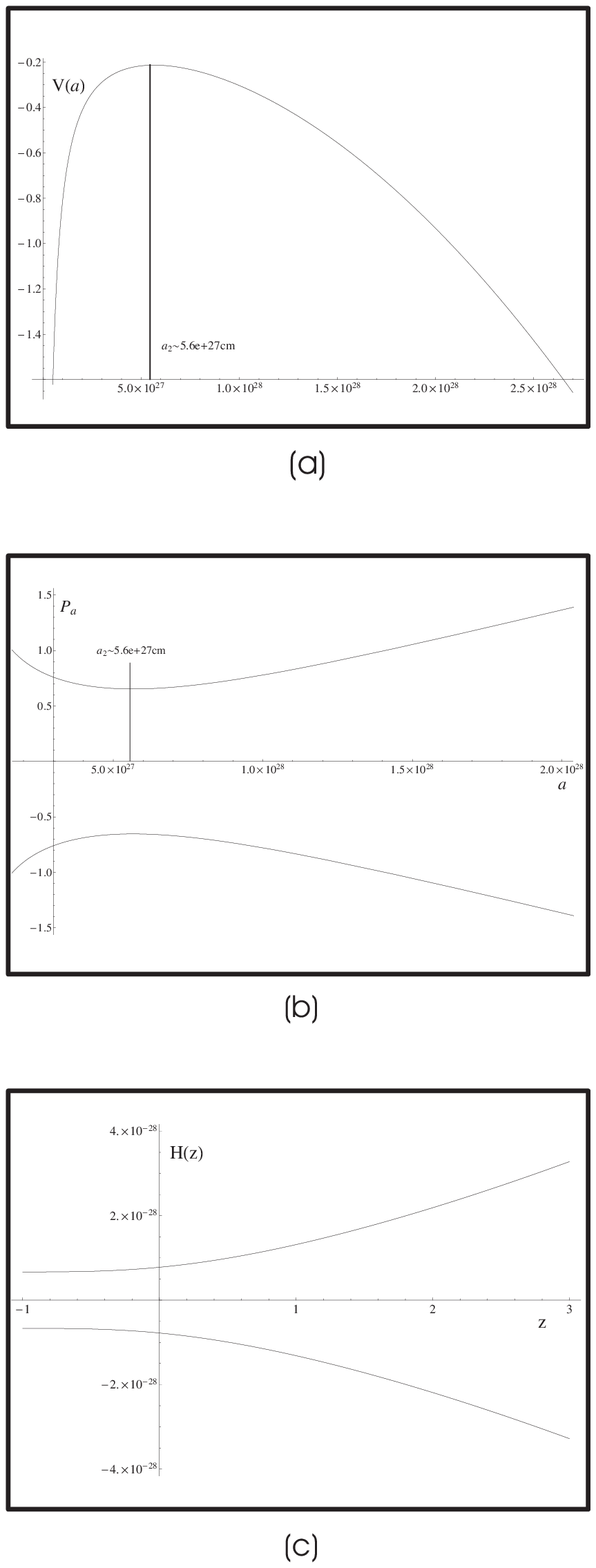}}
\caption{(a)The potential $V(a)$  and the phase space (b) in the region that encompasses the critical point $a_2$ and completes Figs. \ref{Fig. 1}.
This critical point is of the order of $10^{28}~{\rm cm}$, coinciding with the domain of homogeneity and isotropy of our present Universe, regardless of the value of $|\sigma|$.
In (c) we show the behavior of the Hubble factor as function of redshift in a neighborhood of the saddle $a_2$ given the normalization $a_0=1$ and the parameters (\ref{eqrs4}).
The domain $0 \geq z > -1$ of the $H>0$ branch corresponds to the final acceleration phase approaching de Sitter as $z \rightarrow -1$, with $H= {\rm const.}$ at $z=-1$.
}
\label{Fig. 2}
\end{center}
\end{figure}
As observational cosmology asserts, the domain of homogeneity and isotropy of our present Universe is
well accepted for scales around the present horizon, which is given by $a_0\sim 10^{28}~{\rm cm}$ (here, the subscript 0
denotes the present epoch). In this case, we obtain the following observational parameters
\begin{eqnarray}
\label{eqoc1}
~~~\Lambda_4\simeq1.34\times10^{-56} {\rm cm}^{-2}~,~~\\
E_{dust}\simeq2.6\times10^{54}~{\rm g}~,~~~~~~~\\
E_{rad}\simeq4\times10^{78}~{\rm g}~{\rm cm},~~~~~
\end{eqnarray}
where the Hubble radius is fixed to $H_0 \sim 0.77\times 10^{-28}~{\rm cm}^{-1}$.
\par
From Ref. \cite{ijmpd}, the brane tension has a lower bound which corresponds to
$|\sigma|_{min}\sim~10^{22}~{\rm g}~{\rm cm}^{-3}$.
That is, a star with
the Chandrasekhar mass will not form an event horizon if the brane tension is
smaller than $|\sigma|_{min}$. It turns out that
this value furnishes us with a curvature scale
$l_c\equiv 1/\sqrt{R_b}=(\sqrt{a/\ddot{a}})_b\sim{10^{34}}~l_{P}$ at the bounce (where $l_{P}$ is the Planck lenght and
$R_b$ is the Ricci scalar at the bounce). In order to guarantee that $l_c$ at the bounce is not smaller than ${10^3}~l_{P}$, the brane tension must be less than $10^{85}~{\rm g}~{\rm cm}^{-3}$. Therefore, we have the following physical domain (not spoiling the nucleosynthesis) for the brane tension
\begin{eqnarray}
\label{eqocbt}
10^{22}{\rm g}/{\rm cm}^3~\lesssim~|\sigma|~\lesssim~10^{85}~{\rm g}/{\rm cm}^3,
\end{eqnarray}
where we have set $c=1$.
Feeding the Hamiltonian constrain (\ref{eq3.6}) with $|\sigma|_{min}$ and the parameters (\ref{eqoc1}), we obtain that the spatial curvature is
$k\simeq 0.002$ for $|\sigma|\geq|\sigma|_{\min}$.
\par
Considering the lower bound limit for $|\sigma|$, numerical calculations show that the potential $V(a)$ has
always a local minimum at $a_1$ (corresponding to a center) and a local maximum at $a_2$ (corresponding to a saddle). -- cf. Figs. 3(a) and 4(a).
If we increase $|\sigma|$ by four orders of magnitude, we obtain a value of $a_1$ decreased by one order of magnitude.
On the other hand, the local maximum $a_2$ is of the order of $10^{28}~{\rm cm}$ (cf. Fig. 4(a)) for $|\sigma|\geq|\sigma|_{\min}$
(regardless of the value of $|\sigma|$). We exhibit the behavior of $V(a)$ and the phase space $(a,p_a)$ trajectory -- for the parameters (\ref{eqoc1})-(24)
and $|\sigma|\sim|\sigma|_{min}$ -- in Figs. 3(a) and 4(a), and Figs. 3(b) and 4(b), respectively. We should note that Figs. 3 and Figs. 4 display the same potential $V(a)$ and the same universe phase space trajectory in distinct ranges of $a$,
complementing each other.
\par
Given the redshift relation
\begin{eqnarray}
\label{eqrs1}
\frac{a(0)}{a(z)}=1+z,
\end{eqnarray}
equation(\ref{eq4}) can be rewritten as
\begin{eqnarray}
\label{eqrs2}
\nonumber
H^2=H^2_{0}\Big\{\Omega_{0dust}(1+z)^3+\Omega_{0rad}(1+z)^4~~~~~~~~~~\\
\nonumber
+\Omega_{0\Lambda}-\Omega_{0k}(1+z)^2~~~~~~~~\\
-\frac{3H^2_{0}}{16\pi G|\sigma|}(1+z)^6[\Omega_{0dust}+\Omega_{0rad}(1+z)]^2\Big\}
\end{eqnarray}
where
\begin{eqnarray}
\label{eqrs3}
\Omega_{0dust}\equiv\frac{\rho_{0dust}}{\rho^{cr.}_{0}}~,~\Omega_{0rad}\equiv\frac{\rho_{0rad}}{\rho^{cr.}_{0}}~,\\
\Omega_{0\Lambda}\equiv\frac{\rho_{\Lambda}}{\rho^{cr.}_{0}}\equiv\frac{\Lambda}{8\pi G}\frac{1}{\rho^{cr.}_{0}}~,\\
\Omega_{0k}\equiv \frac{k}{a^2_0 H^2_0},
\end{eqnarray}
and $\rho^{cr.}_{0}\equiv 3H^2_0/8\pi G$. By fixing the normalization $a_0=1$, we obtain the following parameters according
to the WMAP 7 year results\cite{wmap}:
\begin{eqnarray}
\label{eqrs4}
\Omega_{0dust}\simeq 0.26~,~\Omega_{0rad}\simeq 10^{-5}~,~
\Omega_{0\Lambda}\simeq 0.73.
\end{eqnarray}
Substituting these parameters in (\ref{eqrs2}), we obtain $\Omega_{0k}\simeq0.004$. In figures 3(c) and 4(c) we show the behavior of the Hubble scale factor
$H$ with respect to the redshift $z$.

It is remarkable that considering the interval of $62$ orders of magnitude of $|\sigma|$ (cf. (\ref{eqocbt})), the trajectory in the phase space of the above {\it observable} universe belongs to region II
of the phase space (cf. Fig. \ref{fig-plane}) corresponding to a one-single-bounce orbit. The part of the trajectory starting
from ($a=a_1,p_{a_{1}}<0$) is an initial acceleration phase that leads the universe through the
bounce and ends in ($a=a_1,p_{a_{1}}>0$), when the universe enters in a long and smooth decelerated expansion phase.
This primordial bouncing accelerated phase does not
correspond to usual inflation, the number of e-folds being
$0.27$. Note, however, that there is no horizon problem in the model.
In fact, before the bounce, due to its cosmological constant
dominated contraction from the infinity past until $a_2$, the particle horizon $d_p$
is given by
\begin{eqnarray}
\label{neq}
d_p=\Big|\bar{a}\int_{\infty}^{\bar{a}}\frac{1}{a\dot{a}}da \Big|\simeq 10^{28} ~{\rm cm},
\end{eqnarray}
if $\bar{a}\geq a_2$. Therefore the particle horizon
is already of
the order of $\Lambda ^{-1/2}$, which is constrained
by present observations to be of the order of the Hubble
radius today. Hence there is no horizon problem for the
scales of cosmological interest. The decelerated expansion Friedmann phase ends in the neighborhood of $a_2$ with a graceful exit to a late de Sitter accelerated phase.
\par
From (\ref{eq1.2.14}) and (\ref{eqoc1}), we see that the parameter $\Lambda_5$ of the model must be adjusted in a very precise way. In fact, $\Lambda_{5}$ must be very close to $8\pi G_N|\sigma|$, which has the minimum value $10^5 {\rm cm}^{-2}$ (see Eq.~(\ref{eqocbt})) and increases as $|\sigma|$ increases, in order to yield the observed value of $\Lambda_4$ given in Eq.~(\ref{eqoc1}).
This is the usual problematic fine-tuning of the cosmological
constant, of at least $60$ orders of magnitude as we have seen above, which the present model, at least in this first approach, does not solve.
It turns out that this is similar to an issue contained in the Randall and Sundrum model \cite{randall}.
In this scenario the brane is embedded in an anti-de Sitter $(4+1)$
spacetime and the fine-tuning relation $\Lambda_5 = - \kappa_5^4\sigma^2/6$ has to be satisfied. It was shown in \cite{durrer} that
the Randall-Sundrum model is unstable under small deviations from this fine-tuning.
As a future investigation we will examine if the same happens in our model.

\section{Conclusions}

In the framework of a Brane World formalism with a timelike extra dimension, we have obtained a homogeneous and isotropic bouncing model
compatible with all observations at the background level. It starts with a de Sitter contraction from the infinity past, experiences a bounce
at very small scales, turning to the usual standard expanding decelerating phases of radiation and matter domination, and a
recent transition to an accelerating expansion. The bounce itself is caused by the appearance of new terms coming from the extra timelike
dimension of the bulk in the $4$-dimensional Friedmann equation, which become important at high curavture scales and avoid the
cosmological singularity, inducing a gravitational repulsion due to the timelike nature of the extra dimension. We have two free parameters:
the brane tension $\sigma$ and the five dimensional cosmological constant $\Lambda_{5}$. The brane tension can assume a wide variety
of values, see Eq.~(\ref{eqocbt}), but $\Lambda_{5}$ must be highly fine-tuned to the value of $\sigma$ in order to yield an effective
$4$-dimensional cosmological constant compatible with observations. Hence the model solves the singularity problem of the standard
cosmological model, together with the horizon and flatness puzzles, but it does not solve the cosmological constant problem.
\par
Our next step will be to perturb the model and investigate the evolution of cosmological perturbations in such cosmological background.
Indeed, our work in progress shows that,  if one imposes an unperturbed de Sitter bulk, a numerical
treatment of linear hydrodynamical perturbation in the universe indicates that the bounce has the effect of substantially enhance the perturbations, nonetheless these perturbations
remain bounded with $\delta \rho/\rho \ll 1$ and $\delta p/p\ll 1$. However, a general analysis of cosmological perturbations
in this scenario demands also a perturbed de Sitter bulk. In this case the $5$-D scalar perturbations will
induce fluctuations of the Weyl tensor projected on the brane which will modify the perturbed field equations.
This is a technical and conceptually involved problem which will be investigated in future publications.

\section*{Acknowledgements}

The authors acknowledge the partial financial support from CNPq/MCTI-Brasil, through a
post-doctoral research grant no. 201907/2011-9 (RM) and
research grant no. 306527/2009-0 (IDS). NPN also would like to thank
CNPq of Brazil for financial support. RM acknowledges the Institute of Cosmology and Gravitation, University of Portsmouth, for their hospitality.
Figures were generated using the Wolfram Mathematica $7$ and MAPLE $13$.

\end{document}